\documentstyle[12pt,fullpage]{article}
\begin{document}
\parskip=10pt
\begin{center}
{\bf Particles and Events in Classical Off-Shell Electrodynamics}

M. C. Land

Department of Communications Engineering  \\
The Center for Technological Education Holon  \\
Affiliated With Tel Aviv University \\
P. O. Box 305, Holon 58102, Israel 

\end{center}
\baselineskip 7mm
\parindent=0cm 

\begin{abstract}
\parindent=0cm 
\parskip=10pt
Despite the many successes of the relativistic quantum theory
developed by \hbox{Horwitz,} et.\ al., certain difficulties
persist in the associated covariant classical mechanics.
In this paper, we explore these difficulties
through an examination of the classical Coulomb problem in the
framework of off-shell electrodynamics.  As the local gauge
theory of a covariant quantum mechanics with evolution parameter
$\tau$, off-shell electrodynamics constitutes a dynamical theory
of spacetime events, interacting through five $\tau$-dependent
pre-Maxwell potentials.  We present a straightforward solution of
the classical equations of motion, for a test event traversing the
field induced by a ``fixed'' event (an event moving uniformly
along the time axis at a fixed point in space).  This solution is
seen to be unsatisfactory, and
reveals the essential difficulties in the formalism
at the classical level.  We then
offer a new model of the particle current --- as a certain distribution of
the event currents on the worldline --- which eliminates these difficulties
and permits
comparison of classical off-shell electrodynamics with the
standard Maxwell theory.  In this model, the ``fixed'' event
induces a Yukawa-type potential, permitting a semi-classical
identification of the pre-Maxwell time scale $\lambda$ with the
inverse mass of the intervening photon.  Numerical solutions to
the equations of motion are compared with the standard Maxwell
solutions, and are seen to coincide when $\lambda >
10^{-6}$ seconds, providing an initial estimate of this
parameter.  It is also demonstrated that the proposed model
provides a natural interpretation for the photon mass cut-off
required for the renormalizability of the off-shell quantum
electrodynamics.

\end{abstract}
\section{Introduction}

In 1973, Horwitz and Piron \cite{H-P} constructed a canonical
formalism for the relativistic classical and quantum mechanics of
many bodies.  To formulate a generalized Hamilton's principle,
they introduced a Poincar\'e invariant evolution parameter $\tau$,
corresponding to the ordering relation among successive events in
spacetime.  This covariant mechanics differs from the `proper time
method' [2 -- 10] in two significant ways: first, by introducing
invariant two-body potentials defined on an unconstrained
eight-dimensional phase space, Horwitz and Piron relaxed the
identification of the parameter $\tau$ with the proper time of
the classical motion, so that particle mass becomes a dynamical
quantity \cite{beyond}.  Second, Horwitz and Piron regard $\tau$
as a true physical time, with the status of the Newtonian time in
non-relativistic mechanics.  Particle worldlines are traced out
through the dynamic evolution of interacting events $x^\mu (\tau)$.
Within this framework, manifestly
covariant solutions have been given for problems in scattering
\cite{scattering}, the bound state \cite{bound} with radiative
transitions \cite{selrul} and Zeeman spectra \cite{zeeman}, and
statistical mechanics \cite{statmech}.

Turning to the question of gauge invariance, Sa'ad, Horwitz, and
Arshansky \cite{saad} constructed a local gauge theory in which
the gauge function depends on $\tau$ as well as the spacetime
coordinates.  This requirement leads to a theory of five gauge
compensation fields, with explicit \hbox{$\tau$-dependence,}
corresponding to electromagnetic modes with continuous mass
spectrum.  In the resulting off-shell electrodynamics, moreover,
particles and gauge fields may exchange mass, even at the
classical level.

A free particle in the quantum mechanics of Horwitz and Piron
satisfies the Stueckelberg equation \cite{H-P,Stueckelberg}
\begin{equation}
i \hbar \partial_{\tau} \psi(x,\tau)
=\frac{1}{2M}p^{\mu}p_{\mu}\psi(x,\tau) \ .
\label{eqn:1}
\end{equation}
The equation
\begin{equation}
(i \hbar \partial_{\tau} + \frac{ e_{0} }{c} \phi) \ \psi(x,\tau)
=\frac{1}{2M}(p^{\mu}-\frac{ e_{0} }{c}a^{\mu})(p_{\mu}-
\frac{ e_{0} }{c}a_{\mu}) \ \psi(x,\tau)
\label{eqn:3}
\end{equation}
is invariant under local gauge transformations of the form
\begin{equation}
\psi(x,\tau) \longrightarrow \left[
\exp \frac{i e_{0} }{\hbar c}
\Lambda (x,\tau)\right]\ \psi (x,\tau)
\label{eqn:2}
\end{equation}
when the compensation fields transform as
\begin{equation}
a_{\mu}(x,\tau) \rightarrow
a_{\mu}(x,\tau)+\partial_{\mu}\Lambda(x,\tau)
\qquad
\phi(x,\tau) \rightarrow \phi(x,\tau) + \partial_{\tau}
\Lambda(x,\tau)  \ .
\label{eqn:4}
\end{equation}
Equation (\ref{eqn:3}) leads to the five dimensional conserved
current
\begin{equation}
\partial_{\mu}j^{\mu}+\partial_{\tau} \rho=0
\label{eqn:5}
\end{equation}
where
\begin{equation}
\rho = \Bigl| \psi(x,\tau) \Bigr| ^{2} \qquad
j^{\mu} = \frac{-i \hbar}{2M} \Bigl\{ \psi^{*}(\partial^{\mu}-i
\frac{ e_{0} }{c}
a^{\mu})\psi - \psi(\partial^{\mu}+i \frac{ e_{0} }{c}
a^{\mu})\psi^{*}
\Bigr\} \ ,
\label{eqn:6}
\end{equation}
so that, $\Bigl| \psi(x,\tau)
\Bigr| ^{2}$ may be interpreted as the probability density at
$\tau$ of finding the event at $x$.  With the summation
convention 
\begin{equation}
\lambda,\mu,\nu = 0,1,2,3 \qquad\qquad {\rm and} \qquad\qquad
\alpha,\beta,\gamma=0,1,2,3,5
\label{eqn:7}
\end{equation}
and the designations
\begin{equation}
x^5 =c \tau \qquad\qquad \partial_5 =\frac1c \partial_\tau \qquad\qquad
j^5 = c \rho
\label{eqn:designations}
\end{equation}
the current conservation law may be written as
$\partial_{\alpha}j^{\alpha}=0$.  Since $\partial_\mu j^\mu
= - \partial_{\tau} \rho \not = 0$, 
we may not identify $j^\mu$ as the source current in
Maxwell's equations.  However, under the boundary conditions
$j^5 \rightarrow 0$, pointwise, as
$\tau \rightarrow \pm \infty$, integration of (\ref{eqn:5})
over $\tau$, leads to $\partial_{\mu}J^{\mu}=0$, where
\begin{equation}
J^{\mu}(x)=\int_{-\infty}^{\infty} d\tau \ j^{\mu}(x,\tau) \ .
\label{eqn:8}
\end{equation}
This integration has been called concatenation \cite{concat} and
links the event current $j^\mu$ with the particle current
$J^\mu$ whose support covers the entire worldline.  The quantum mechanical
potential theory with $a_\mu =0$ and $- \frac{ e_{0} }{c} \phi =
V(\sqrt{x^\mu x_\mu})$ has been solved for the standard bound
state \cite{bound} and scattering \cite{scattering} problems.

The classical mechanics associated with this theory is obtained
by transforming the Hamiltonian found from (\ref{eqn:3}) to a
classical Lagrangian \cite{emlf}, and including the gauge invariant
kinetic term for the fields proposed by Sa'ad, et.\ al.\ 
\cite{saad}:
\begin{equation}
L = \frac12 M \dot x^\mu \dot x_\mu + \frac{ e_{0} }{c}
\dot x^\mu a_\mu + e_{0} \phi
-\frac{\lambda}{4c} f^{\alpha\beta} f_{\alpha\beta}
= \frac12 M \dot x^\mu \dot x_\mu + \frac{ e_{0} }{c}
\dot x^\alpha a_\alpha
-\frac{\lambda}{4c} f^{\alpha\beta} f_{\alpha\beta} \ .
\label{eqn:14}
\end{equation}
In (\ref{eqn:14}), we have used $\dot x^5 = c$ and introduced
$a_5 = \frac1c \ \phi$.
The gauge invariant quantity $f_{\alpha\beta}$ is defined by
\begin{equation}
f_{\alpha\beta} =\partial_\alpha a_\beta - \partial_\beta
a_\alpha \ .
\label{eqn:13}
\end{equation}
The classical Lorentz force \cite{emlf} found by variation of
(\ref{eqn:14}) with respect to $x^\mu$, is given by
\begin{equation}
M \: \ddot x^\mu = \frac{ e_{0} }{c}
 \ f^\mu_{\:\;\;\alpha}(x,\tau) \, \dot
x^\alpha \qquad \qquad \frac{d}{d\tau} (-\frac{1}{2} M \dot x^2) =
e_0 \ f_{5\alpha} \dot x^\alpha \ .
\label{eqn:12}
\end{equation}
From the second of equation (\ref{eqn:12}), one sees that the
exchange of mass between particles and fields is induced by
$f_{5\alpha}$, and thus made possible the fifth potential $a_5$ and
the $\tau$-dependence of $a_\mu$.  Since particle mass is not
separately conserved, pair annihilation is classically permitted;
the total mass-energy-momentum of the events and fields is
conserved \cite{emlf}, however.

In formally raising the index $\beta=5$ in
$f^{\mu 5} = \partial^\mu a^5 - \partial^5 a^\mu$, Sa'ad et.\ al.\
argue that the action suggests a higher symmetry containing
O(3,1) as a subgroup, that is, either O(4,1) or O(3,2).  They
wrote the metric for the field as
\begin{equation}
g^{\alpha\beta} = {\rm diag}(-1,1,1,1,\sigma) \ ,
\label{eqn:15}
\end{equation}
where $\sigma = \pm 1$, depending on the higher symmetry.
Variation of (\ref{eqn:14}) with respect to $a_\alpha$ yields
\begin{equation}
\partial_{\beta} f^{\alpha \beta}
=\frac{e_{0}}{\lambda c}j^{\alpha}=\frac{e}{c} \ j^{\alpha}
\qquad\qquad
\epsilon^{\alpha \beta \gamma \delta
\epsilon}\partial_{\alpha}f_{\beta \gamma}=0
\label{eqn:16}
\end{equation}
where $e_0/\lambda$ is identified as the dimensionless charge
$e$, and the current $j^\alpha$ associated with an event
$X^\alpha = \Bigl(X^\mu(\tau),c\tau\Bigr)$
is given by
\begin{equation}
j^\alpha (x,\tau) = c \ \frac{dX^\alpha}{d\tau} \delta^4
\Bigl(x^\mu-X^\mu(\tau)\Bigr) \ .
\label{eqn:17}
\end{equation}
At the quantum level, the current is given by (\ref{eqn:6}).

In analogy to the concatenation of the current in (\ref{eqn:8}),
we see that under the boundary conditions $f^{5\mu} \rightarrow
0 $, pointwise as $\tau \rightarrow \pm \infty$, we recover Maxwell's
equations as
\begin{equation}
\partial_{\nu}F^{\mu \nu}=\frac{e}{c} J^{\mu} \qquad\qquad
\epsilon^{\mu \nu \rho \lambda }\partial_{\mu}F_{\nu \rho}=0
\label{eqn:18}
\end{equation}
where
\begin{equation}
F^{\mu \nu}(x)=\int_{-\infty}^{\infty} d\tau f^{\mu \nu}(x,\tau)
\qquad\qquad {\rm and} \qquad\qquad
A^{\mu}(x)=\int_{-\infty}^{\infty} d\tau \ a^{\mu}(x,\tau) \ .
\label{eqn:19}
\end{equation}
Therefore, $a^{\alpha}(x,\tau)$ has been called the pre-Maxwell
field.  Since $e_0 a_\mu$ and $e A_\mu$ must have the same
dimensions, it follows from (\ref{eqn:19}) that $\lambda$ (and
hence $e_0 = \lambda e$) must have dimensions of time.  Although
the parameter $\lambda$ does not appear in the field equations
(\ref{eqn:16}), it does appear in the Lorentz force
(\ref{eqn:12}) through $e_0$.  The presense of this dimensional
parameter in the equations of motion is a characteristic problem
in the classical formalism.

The physical Lorentz covariance of the current $j^\alpha$ breaks
the formal higher symmetry --- O(3,2) or O(4,1) --- of the free field
equations to O(3,1).  Nevertheless, the wave equation
\begin{equation}
\partial_{\alpha}\partial^{\alpha} \ a^{\beta}=
(\partial_{\mu}\partial^{\mu}+\partial_{\tau}
\partial^{\tau}) \ a^{\beta}=
(\partial_{\mu}\partial^{\mu}+ \sigma \; \partial_{\tau}^2) \ 
a^{\beta} = -\frac{e}{c} \ j^{\beta} \ ,
\label{eqn:20}
\end{equation}
reflects the causal properties of the higher symmetry through the
operator on the left hand side.  The classical Green's
function for (\ref{eqn:20}), defined through
\begin{equation}
\partial_{\alpha}\partial^{\alpha}G(x,x^5) = - c \delta^4 (x,x^5)
\ ,
\label{eqn:21}
\end{equation}
is given by \cite{green}
\begin{equation}
G(x,x^5) = - {c\over{4\pi}}
\delta(x^2)\delta(x^5) -
{c\over{2\pi^2}}{\partial\over{\partial{x^2}}} \
{{\theta(-\sigma g_{\alpha\beta}
x^\alpha x^\beta)}\over{\sqrt{-\sigma g_{\alpha \beta}
x^\alpha x^\beta}}} \ .
\label{eqn:22}
\end{equation}
It follows from (\ref{eqn:20}) and (\ref{eqn:21}) that the
potential induced by a known current is given by
\begin{eqnarray}
a^\beta (x,\tau) &=& \frac{e}{c} \int d^4x' dx^{5\prime} \ 
\frac1c \ G(x-x',x^5 - x^{5\prime}) \ j^\beta (x',c\tau')
\nonumber \\
&=& \frac{e}{c} \int d^4x' d\tau  \ G(x-x',c\tau - c\tau')
 \ j^\beta (x',c\tau')
 \ .
\label{eqn:23}
\end{eqnarray}
Under concatenation, the first term of (\ref{eqn:22}) becomes
the Maxwell Green's function
\begin{equation}
D(x) = - {c\over{4\pi}} \int d\tau \ \delta(x^2)\delta(c \tau) 
= - {1\over{4\pi}} \ \delta(x^2) \ ,
\label{eqn:24}
\end{equation}
while the second term --- which induces spacelike or timelike
correlations \cite{green}, depending on the signature
$\sigma$ --- vanishes.
This concatenation guarantees that the Maxwell potential is
related to the Maxwell current in the usual manner:
\begin{eqnarray}
A^\mu (x) &=& \int d\tau \ a^\mu (x,\tau)
\nonumber \\
&=& -\frac{e}{c} \int d\tau  \int d^4x' d\tau' \ 
G(x-x',c\tau-c\tau') \
j^\mu (x',c\tau')
\nonumber \\
&=& -\frac{e}{c} \int d^4x' d\tau' \ 
\Bigl[ \int d\tau G(x-x',c\tau-c\tau')
\Bigr] \ j^\mu (x',c\tau')
\nonumber \\
&=& -\frac{e}{c} \int d^4x' D(x-x') \ J^\mu (x') \ .
\label{eqn:25}
\end{eqnarray}
Therefore, we will refer to the Maxwell and the correlation terms
of the Green's function and the induced potentials.

The off-shell quantum electrodynamics, associated with the action
\begin{equation}
{\rm S} = \int d^4 x d\tau \left\{ 
\psi^* (i\partial_\tau + e_0 a_5 ) \psi - 
\frac{1}{2M} \psi^* (-i\partial_\mu - e_0 a_\mu )
    (-i\partial^\mu - e_0 a^\mu )\psi 
- \frac{\lambda}{4} f_{\alpha\beta}f^{\alpha\beta} \right\} \ ,
\label{eqn:action}
\end{equation}
(here, $\hbar = c =1$) has been worked out \cite{qed}.
Manifestly covariant
quantization has been given canonically \cite{shnerb,qed} and in
path integral form \cite{jaime,qed}, and the perturbation theory
developed \cite{qed}.  The Feynman rules have been used to
calculate the
scattering cross section for two identical particles; this cross
section reduces to the standard Klein-Gordon expression when no
mass exchange is permitted \cite{qed}.  For any non-zero mass
exchange, the forward and reverse poles each split into two and
move away from the 0 and 180 degree directions, making the total
cross section finite.  When the photon mass spectrum is cut off,
the off-shell quantum electrodynamics is counter-term
renormalizable; without the cut-off, the mass integration in the
photon loops cannot be controlled.  We will see below that this
cut-off has a natural interpretation in terms of the classical
model presented below.

We now turn to the Coulomb problem in the framework of the
classical off-shell electromagnetic theory introduced above.  In
Section 2, we set up the classical equations of motion for a test
event, scattering in the field of a source event which moves uniformly
along the time axis, and present a straightforward solution.  This solution
is seen to be plainly unsatisfactory, and suggests an alternative
approach to the particle current.  In Section
3, we propose a new model for the relationship between events and
particles, which leads to a finite-mass Yukawa-type potential.
We solve the resulting equations of motion, and compare the
results the standard Maxwell solutions.

This paper is dedicated to Professor L.\ P.\ Horwitz, whose
deep contributions to our understanding of relativistic dynamics
have re-opened this subject for a generation of students.

\section{The Coulomb Problem}

We begin by studying the motion of a test event in the field
produced by a source event, moving uniformly along the time axis.
As in classical Rutherford scattering, we hold the source event
``fixed'' on its time axis, neglecting the field of the scattered
event, and the radiation field that its acceleration would induce.

According to equation (\ref{eqn:17}),
the current associated with a source event moving uniformly along
the time axis,
\begin{equation}
X^0 (\tau) = ct = c\tau \qquad\qquad \vec{X}(\tau) = 0
\label{eqn:26}
\end{equation}
is given by
\begin{eqnarray}
j^0 (x,\tau) &=& c \frac{dX^0}{d\tau} \delta(x^0 - c\tau)
\delta^3(\vec{x}) = c \ \delta(t - \tau) \delta^3(\vec{x})
\nonumber \\
\vec{j} (x,\tau) &=& 0
\nonumber \\
j^5 (x,\tau) &=& c \frac{d(c\tau)}{d\tau} \delta(x^0 - c\tau)
\delta^3(\vec{x}) = j^0
(x,\tau) \ .
\label{eqn:27}
\end{eqnarray}
Therefore, from (\ref{eqn:23}) we have
\begin{equation}
a^5 (x,\tau) = a^0 (x,\tau) \qquad \qquad \vec{a} (x,\tau) = 0
\ ,
\label{eqn:28}
\end{equation}
so that the only non-zero components of $f^{\alpha\beta}$ are
\begin{eqnarray}
f^{0k} &=& \partial^0 a^k - \partial^k a^0 = - \partial^k a^0
\nonumber \\
f^{5k} &=& \partial^5 a^k - \partial^k a^5 = - \partial^k a^0
\nonumber \\
f^{05} &=& \partial^0 a^5 - \partial^5 a^0 =
- \frac{1}{c} \left( \partial_0 a^0 + \sigma \partial_\tau a^0 \right) \ .
\label{eqn:29}
\end{eqnarray}
The independent components of the Lorentz force may now be
written as
\begin{eqnarray}
M \ \ddot x^0 &=& \frac{e_0}{c} \ f^{0\alpha} \dot x_\alpha
\nonumber \\
&=& \lambda \frac{e}{c} \ (f^{0k} \dot x_k + f^{05} \dot x_5)
\nonumber \\
&=& - \lambda \frac{e}{c} \ \Bigl[ (\partial_k a^0) \dot x^k +
\sigma \frac1c \ (\partial_0 + \sigma \partial_\tau) \ a^0 \Bigr]
\label{eqn:30}
\end{eqnarray}
and
\begin{eqnarray}
M \ \ddot x^k &=& \frac{e_0}{c} \ f^{k\alpha} \dot x_\alpha
\nonumber \\
&=& \lambda \frac{e}{c} \ (f^{k0} \dot x_k + f^{k5} \dot x_5)
\nonumber \\
&=& - \lambda \frac{e}{c} \ (\partial_k a^0)(\dot x^0 - \sigma c) \ ,
\label{eqn:31}
\end{eqnarray}
where we used the antisymmetry of $f^{\alpha\beta}$ and $\dot x_5
= \sigma c$.  Since at low energies, $\dot x^0 \sim c$, we notice
that the choice of $\sigma =1$ in (\ref{eqn:31}) rules out
identification with the standard Maxwell equations of motion.  We
therefore choose $\sigma = -1$ and study only the case of broken
O(3,2) symmetry.  With this choice,
\begin{eqnarray}
M \ \ddot x^0 &=& - \lambda \frac{e}{c} \ 
\Bigl[ (\partial_k a^0) \dot x^k -
\frac1c \ (\partial_t - \partial_\tau) \ a^0 \Bigr]
\label{eqn:32}
\\
M \ \ddot x^k &=&  - \lambda e \ 
(\partial_k a^0)(\dot t +1)  \ .
\label{eqn:33}
\end{eqnarray}
Since $\lambda$ does not appear in equations (\ref{eqn:23})
or (\ref{eqn:27}), the coupling of the induced
field to the test event will evidently depend on this parameter.

We proceed to calculate the potential $a^0 (x,\tau)$ from the Green's
function (\ref{eqn:22}) and the current (\ref{eqn:27}).  The
Maxwell part of the potential is given by
\begin{eqnarray}
a^0 (x,\tau) &=& -\frac{e}{c} \int d^4x' d\tau' \Bigl[
\frac{c}{4\pi}  
 \ \delta \Bigl( \ (x-x')^2 \ \Bigr) \ 
 \delta(c\tau-c\tau') \Bigr] \ 
\Bigl[c \ \delta^3 (\vec{x}') \ \delta(t' - \tau') \Bigr]
\nonumber \\
&=& -\frac{e}{4\pi c} \int d^3x' d(ct') \ 
\delta\Bigl( (\vec{x}-\vec{x}')^2 - (x^0 - x^{0\prime})^2 \Bigr) \ 
\Bigl[c \ \delta^3 (\vec{x}') \ \delta(t' - \tau) \Bigr]
\nonumber \\
&=& -\frac{e c}{4\pi} \delta\Bigl( R^2 - c^2(t -\tau)^2 \Bigr) 
\nonumber \\
&=& -\frac{e}{4\pi R} \ \frac12 \Bigl[ \delta(t - \tau - R/c) +
\delta(t - \tau + R/c) \Bigr] \ ,
\label{eqn:34}
\end{eqnarray}
where $R=|\vec{x}|$.  As required for the Maxwell part,
\begin{equation}
A^0 (x) = \int d\tau a^0 (x,\tau) =  - \frac{e}{4\pi R} \ ,
\label{eqn:35}
\end{equation}
the concatenated potential has the form of the Coulomb potential
due to a ``fixed'' source.
The second part of the pre-Maxwell potential is given by
\begin{eqnarray}
a_{\rm correlation} (x,\tau) &=& \frac{e}{c} \ 
\frac{c}{2\pi^2} \int d^4x' d\tau'
\left\{ \frac{\partial}{\partial (x-x')^2 }  \
\frac{\theta\Bigl( (x-x')^2 - c^2(\tau - \tau' )^2 \Bigr)}
{ \sqrt{(x-x')^2 - c^2(\tau - \tau' )^2} } \right\} \times
\nonumber \\&&\mbox{\qquad}
c \ \delta^3 (\vec{x}') \ \delta(t' - \tau')
\nonumber \\
&=& \frac{ec^2}{2\pi^2} \int d\tau' \ 
\frac{\partial}{\partial R^2 }  \ 
\frac{\theta\Bigl( R^2 - c^2(t - \tau' )^2
- c^2(\tau - \tau' )^2 \Bigr)}
{ \sqrt{R^2 - c^2(t - \tau' )^2 - c^2(\tau - \tau' )^2  } }
\nonumber \\
&=& \frac{ec^2}{2\pi^2} \frac{\partial}{\partial R^2 }  \ 
\int d\tau' \ 
\frac{\theta\Bigl( R^2 - c^2(t - \tau' )^2
- c^2(\tau - \tau' )^2 \Bigr)}
{ \sqrt{R^2 - c^2(t - \tau' )^2 - c^2(\tau - \tau' )^2 } } \ .
\label{eqn:36}
\end{eqnarray}
Introducing the change of variables $u=t-\tau'$ and defining
$\alpha = t-\tau $, equation (\ref{eqn:36}) becomes
\begin{equation}
a_{\rm correlation} (x,\tau) =
\frac{ec^2}{2\pi^2} \frac{\partial}{\partial R^2 }  \ 
\int du \ \frac{\theta\Bigl( R^2 - c^2u^2
- c^2(\alpha - u )^2 \Bigr)}
{ \sqrt{R^2 - c^2u^2 - c^2(\alpha - u )^2 } } \ .
\label{eqn:37}
\end{equation}
The $\theta$-function in the integral restricts the limits of
integration to the region between the roots $u_{\pm}$ of the
quadratic
\begin{equation}
\rho(u) = R^2 - c^2u^2 - c^2(\alpha - u )^2 = R^2 - c^2\alpha^2 
+ 2c^2\alpha u - 2 c^2u^2 \ ,
\label{eqn:38}
\end{equation}
which are found to be
\begin{equation}
u_{\pm} = \frac12 \left[ \alpha \pm \sqrt{\frac{2R^2}{c^2}
- \alpha^2} \ \right]
\ .
\label{eqn:39}
\end{equation}
So for $2R^2/c^2 < \alpha^2$, there are no roots and $\theta
(\rho(u))$ will vanish identically.  Therefore,
\begin{equation}
a_{\rm correlation} (x,\tau) =
\frac{ec^2}{2\pi^2} \frac{\partial}{\partial R^2 }  \ 
\left\{ \begin{array}{ll}
\int_{u_{-}}^{u_{+}} du \frac{1}{\sqrt{\rho(u)}} &
\mbox{if $\sqrt{2} R/c < \alpha < \sqrt{2} R/c$}\\
0 & \mbox{otherwise} 
         \end{array}
\right.
\label{eqn:42}
\end{equation}
Using
\begin{equation}
\int du \frac{1}{\sqrt{A+Bu+Cu^2}} = \frac{-1}{\sqrt{-C}} \ 
\sin^{-1} \left( \frac{2Cu+B}{\sqrt{B^2-4AC}} \right)
\label{eqn:43}
\end{equation}
we find
\begin{equation}
a_{\rm correlation} (x,\tau) =
\frac{ec^2}{2\pi^2} \frac{\partial}{\partial R^2 }  \ 
\left\{ \begin{array}{ll}
\frac{-1}{\sqrt{2}} \ 
\left. \sin^{-1} \left( \frac{2(-2) \ u + 2\alpha}
{ \sqrt{4\alpha^2 + 4 (R^2/c^2-\alpha^2) (-2) } }\right) 
\right|_{u_{-}}^{u_{+}} &
\mbox{if $\sqrt{2} R/c < \alpha < \sqrt{2} R/c$}\\
0 & \mbox{otherwise} 
         \end{array}
\right.
\label{eqn:44}
\end{equation}
and inserting (\ref{eqn:39}) we obtain
\begin{equation}
\sin^{-1} \left( \frac{2(-2) \ u_{\pm} + 2\alpha}
{ \sqrt{4\alpha^2 + 4 (R^2/c^2-\alpha^2) (-2) } }\right)
= \sin^{-1} \left( \frac{-( \alpha \pm \sqrt{2 R^2c^2-\alpha^2 } )
+ \alpha}
{ \sqrt{2 R^2c^2-\alpha^2 } }\right) = \sin^{-1} (\mp 1) \ .
\label{eqn:45}
\end{equation}
This last expression is independent of $R$, and so from
(\ref{eqn:44}) we see that
$a_{\rm correlation}$ will vanish.  It may be shown that the
correlation term will only contribute when the inducing charge
undergoes acceleration.  This contribution will be treated
elsewhere.

Having obtained the potential induced by the source current, we
may calculate the field strengths and write the equations of
motion for a test event moving in this field.
The support of the potentials is restricted, by the
delta-functions they contain, to the light cone of
the source event and so the
test event moving in the induced field will be free except when
its trajectory satisfies
\begin{equation}
t - \tau - R/c = 0 \ .
\label{eqn:46}
\end{equation}
We write the initial motion of the test event as
\begin{equation}
x = u \tau + s
\label{eqn:47}
\end{equation}
and choose the constants of the motion such that,
\begin{equation}
s = (s_t, s_x , s_y , 0) \qquad \mbox{and}
\qquad u = (u_t , u_x , 0 , 0)
\label{eqn:48}
\end{equation}
which confine the motion to the $t-x$ plane, with impact
parameter $s_y$.  To facilitate comparison with the
non-relativistic case, we take the velocity
\begin{equation}
\vec{v} = \frac{d\vec{x}}{dt} = \frac{d\vec{x} / d\tau}{dt / d\tau}
\label{eqn:49}
\end{equation}
such that $|\vec{v}| \ll c$, and so
$\frac1c u_t  = dt / d\tau \sim 1$.  Then
the interaction will occur at $\tau^*$, given by
\begin{equation}
0 = t - \tau^* - R/c = \frac1c u_t \tau^* + \frac1c s_t -\tau^* -
\frac1c | \vec{u} \tau^* + \vec{s}|
= \frac1c s_t - \frac1c | \vec{u} \tau^* + \vec{s}| + o(v/c) \ ,
\label{eqn:50}
\end{equation}
and so
\begin{equation}
s_t = \sqrt{(s_x + u_x \tau^*)^2 +s_y^2} \ .
\label{eqn:51}
\end{equation}
Solving (\ref{eqn:51}), we find
\begin{equation}
\tau^* = \frac{1}{u_x} \left(- s_x + \sqrt{s_t^2 - s^2_y} \ \right)
\label{eqn:52}
\end{equation}
so that
\begin{equation}
R^* = R(\tau^*) = s_t \qquad x^* = x(\tau^*) = \sqrt{s_t^2 - s^2_y}
\qquad y^* = y(\tau^*) = s_y
\label{eqn:53}
\end{equation}
and
\begin{equation}
\hat{x}^* = \hat{x}(\tau^*) = \frac{1}{R^*} \vec{x}^* =
\frac{1}{s_t} \left(\sqrt{s_t^2 - s^2_y} , s_y , 0 \right) \ .
\label{eqn:54}
\end{equation}
In the neighborhood of $\tau^*$  we have
\begin{equation}
M \ddot x^k = - \lambda e \ \partial_k \left[ -\frac{e}{8\pi R}
\delta (t-\tau-R/c) \right] (\dot t +1)
\simeq - \lambda e \ \partial_k \left[ -\frac{e}{8\pi R}
\delta (t-\tau-R/c) \right] 2
\label{eqn:55}
\end{equation}
which may be integrated over $\tau$ to obtain
\begin{equation}
\dot{\vec{x}} (\tau^* + \epsilon) - \dot{\vec{x}} (\tau^* - \epsilon) =
\frac{\lambda e^2}{4\pi M} \nabla
\int^{\tau^* + \epsilon}_{\tau^* - \epsilon} d\tau
\frac1R \delta (t-\tau-R/c) = - \frac{\lambda e^2}{4\pi M}
\frac{1}{(R^*)^2} \hat{x}^* \ .
\label{eqn:56}
\end{equation}
Similarly,
\begin{equation}
\dot t (\tau^* + \epsilon) - \dot t (\tau^* - \epsilon) =
- \frac{\lambda e^2}{Mc} \frac{1}{8\pi(R^*)^2} \ 
{\hat{x}}^* \cdot {\dot{\vec{x}}}^* = o(v/c) \sim 0  \ .
\label{eqn:57}
\end{equation}
Since $\vec{x} (\tau^* - \epsilon) = \vec{u}$, the final velocity
will be
\begin{equation}
\vec{u}' = \dot{\vec{x}} (\tau^* + \epsilon) =
\vec{u} - \frac{\lambda e^2}{4\pi M} \ \frac{1}{(s_t)^3} \ 
\left(\sqrt{s_t^2 - s^2_y} , s_y , 0\right) \ .
\label{eqn:58}
\end{equation}
Imposing conservation of energy, $\vec{u}^{\prime 2} =
\vec{u}^2$,
\begin{equation}
\vec{u}^2 = \vec{u}^2 +
\left(\frac{\lambda e^2}{4\pi M} \ \frac{1}{(s_t)^3}\right)^2
- 2 \frac{\lambda e^2}{4\pi M} \ \frac{1}{(s_t)^3}
 \ \left(\sqrt{s_t^2 - s^2_y} , s_y , 0\right) \cdot (u_x , 0 , 0)
\label{eqn:59}
\end{equation}
leads to the requirement
\begin{equation}
0 = \frac{\lambda e^2}{4\pi M} - 2 u_x s_t \sqrt{s_t^2 - s^2_y}
 \ .
\label{eqn:60}
\end{equation}
The scattering angle is given by
\begin{equation}
\cos \theta = \frac{u_x^\prime}{|\vec{u}|} =
\frac{u_x - \frac{\lambda e^2}{4\pi M} \ \frac{1}{(s_t)^3} \sqrt{s_t^2 - s^2_y}}{u_x}
\label{eqn:61}
\end{equation}
and using
\begin{equation}
\cot \frac{\theta}{2} = \sqrt{ \frac{1+\cos\theta}{1-\cos\theta} }
\label{eqn:62}
\end{equation}
and (\ref{eqn:60}) we obtain
\begin{equation}
\cot \frac{\theta}{2} = \frac{s_y}{\sqrt{s_t^2 - s^2_y}} \ .
\label{eqn:63}
\end{equation}
This may be compared with the asymptotic scattering angle for
non-relativistic Rutherford scattering, given by \cite{goldstein}
\begin{equation}
\cot \frac{\theta}{2} = \frac{2Es_y}{e^2/ 4\pi} = \frac{4 \pi M
u^2 s_y}{e^2} \ .
\label{eqn:64}
\end{equation}
Requiring equality of (\ref{eqn:63}) and (\ref{eqn:64}) leads to
\begin{equation}
\sqrt{s_t^2 - s^2_y} = \frac{e^2}{4 \pi M u^2} \ ,
\label{eqn:65}
\end{equation}
which fixes a particular value for $s_t$.  Notice that under this
equality, (\ref{eqn:60}) becomes
\begin{equation}
\lambda = \frac{4 \pi M u s_t}{e^2} \ \sqrt{s_t^2 - s^2_y} =
\frac{4 \pi M u s_t}{e^2} \ \frac{e^2}{4 \pi M u^2} =
\frac{s_t}{u} = \frac{R(\tau^*)}{u}
\label{eqn:66}
\end{equation}
so that the scattering angle will agree with the asymptotic
value obtained in the standard Maxwell
theory only if $\lambda$ is equal to the distance at the time of
interaction divided by the incoming speed.

The solution presented above is unsatisfactory for a number of
reasons, aside from the essential discrepancy between the
piecewise linear solution found here and the smooth acceleration
expected in the Coulomb problem.  Notice first that from
(\ref{eqn:52}) we will find no solution for $\tau^*$ if
$s_t < s_y$, meaning that a test event with these constants of
motion will pass the source event without interaction.  In
particular, two events moving together along the time axis, at
separation $R$ but no relative time offset at $\tau=0$, will
experience no Coulomb force.  The second problem is that agreement
with the standard (asymptotic) scattering angle in Coulomb
scattering can only be obtained if the parameter $\lambda$
depends on the constants of motion of the experiment in a
particular way.

It is the dependence of the scattering angle on $s_t$, the
value of the time coordinate at $\tau = 0$, which points to
the essential difficulty in off-shell electromagnetic theory at
the classical level.  In the standard proper-time formulation of
Maxwell electrodynamics \cite{Schwinger,S-F}, there is no particular
significance to
the parametrization of the worldline, because neither the current
$J^\mu (x)$, nor the potential $A^\mu (x)$ it induces, depends on
$\tau$ --- their support covers the entire worldline, and the
parameter may be assigned arbitrarily \cite{beyond}.

In pre-Maxwell electrodynamics, however, particle worldlines are
traced-out according to the $\tau$-evolution of events, and the
parametrization of the worldlines determines which events undergo
interaction.  Thus, the Green's function (\ref{eqn:22})
determines an interaction which depends upon the spacetime
separation $x^\mu (\tau) - x^{\prime\mu} (\tau)$ of the
interacting events at a given $\tau$ through
\begin{equation}
j^\mu (x,\tau) a_\mu (x,\tau)  =
\int d^4x' d\tau' \ j^\mu (x,\tau) \ G(x-x',\tau-\tau')
\ j^\mu (x',\tau') \ .
\label{eqn:pre-max_int}
\end{equation}
We refer to this mutual parametrization at $\tau$ as
the $\tau$-synchronization (for a freely moving event
of equation (\ref{eqn:47}), this choice is
equivalent to choosing $s^\mu$, its spacetime position at $\tau=0$).
Since the spacetime separation $x^\mu (\tau) - x^{\prime\mu}
(\tau)$, and the corresponding Lorentz force, depend upon the
$\tau$-synchronization of the worldlines, a change in the
$\tau$-parameterization of just one worldline will change the
course of the
interaction.  This dependence was seen in the scattering problem
discussed above, in which the scattering angle was seen to depend on the
relative values of $s_t$ and the impact parameter $s_y$.  In the
absense of interaction, the \hbox{$\tau$-parameterization} of the
worldline has no significance, because the observed Maxwell current
$J^\mu (x)$ is found
by concatenating the pre-Maxwell current $j^\mu (x,\tau)$, and it
retains no information regarding the
\hbox{$\tau$-parameterization}.
Under interaction, however, the \hbox{$\tau$-parameterization} can be made
in an infinite number of dynamically inequivalent ways.

Arshansky, Horwitz, and Lavie \cite{concat} have argued
that measurements made at a spacetime point $x^\mu$ do not take
place at a definite $\tau$, but rather concatenate all events ---
occurring at various values of $\tau$ --- which may causally contribute
to the event at $x^\mu$.  From this point of view, the initial
$\tau$-synchronization of events in a scattering experiment,
and hence $s^\mu$ in the present case,
cannot be precisely measured, even at the classical
level, and may be associated with a fundamental uncertainty. 
In the quantum regime, since states are defined with sharp
asymptotic masses, the initial conditions in $\tau$ are completely
undetermined because of the uncertainty relation between $\tau$
and mass \cite{uncert}.  Therefore, results in relativistic
quantum mechanics and off-shell quantum electrodynamics do not
suffer this dependence.  The success of the formalism in the
quantum regime suggests that the formulation of the classical
equations of motion be modified to take account of this
uncertainty.

\section{Particles as Distributions of Events}

In this section, we present a model which overcomes many of the
difficulties presented in the previous chapter.  We wish to
incorporate in the description of interactions among classical particles an
uncertainty in their $\tau$-synchronization,
and we take as our starting point the observation that the
Maxwell current $J^\mu (x)$ determined by measurement devices
\cite{concat} is insensitive to this uncertainty.  To see this,
we take a normalized distribution \hbox{function $\varphi(\alpha)$},
\begin{equation}
\int_{-\infty}^{\infty} d\alpha \ \varphi(\alpha) =1 \ .
\label{eqn:67}
\end{equation}
and replace
the event current $j^\mu (x,\tau)$, given in (\ref{eqn:17}) for
the sharply defined event $x^\mu(\tau)$, with the current
\begin{eqnarray}
j^\beta_\varphi (x,\tau) &=&
\int_{-\infty}^{\infty} d\alpha \ \varphi(\alpha) \ 
j^\beta (x,\tau-\alpha)
\nonumber \\
&=& c \int_{-\infty}^{\infty} d\alpha \ \varphi(\alpha) \ 
\dot X^\beta (\tau-\alpha)  \ \delta^4
\Bigl(x-X(\tau-\alpha)\Bigr) \ .
\label{eqn:68}
\end{eqnarray}
Since the concatenation integral may be shifted by
$\tau \longrightarrow \tau' = \tau-\alpha$,
the Maxwell current found from (\ref{eqn:68}) will be
\begin{equation}
J^\mu_\varphi (x) = \int d\tau d\alpha \ \varphi(\alpha)
\ j^\mu(x,\tau-\alpha)
= \int d\alpha  \ \varphi(\alpha) \times \int
d\tau' \ j^\mu(x,\tau') = J^\mu (x) \ ,
\label{eqn:70}
\end{equation}
identical to that for the sharply defined event.

Thus, we may model the particle current as a collection of event
currents whose
initial conditions in $\tau$ are given by a smooth distribution.
The microscopic dynamics consists of events interacting through
the pre-Maxwell fields, but the pre-Maxwell current which induces
those fields will be a superposition of the individual event
currents.

In the particular case discussed in the previous section, the
current associated with a particle moving uniformly along the
time axis is given by
\begin{equation}
j^0_\varphi (x,\tau) = \int_{-\infty}^{\infty} d\alpha \ \varphi(\alpha) \ 
j^0 (x,\tau-\alpha) = c \ 
\delta^3(\vec{x}) \ \varphi(t-\tau) \ .
\label{eqn:72}
\end{equation}
The pre-Maxwell potential induced by this current is then
\begin{eqnarray}
a^0_\varphi (x,\tau) &=& -\frac{e}{4\pi} \int d^4x' d\tau' \ 
\delta^3 (\vec{x}') \ \varphi(t' - \tau') \ 
\delta\Bigl( \ (x-x')^2 \ \Bigr) \delta(\tau-\tau')
\nonumber \\
&=& -\frac{e}{4\pi R} \ \frac12 \Bigl[ \varphi(t - \tau - R/c) +
\varphi(t - \tau + R/c) \Bigr] \ ,
\label{eqn:73}
\end{eqnarray}
so that
\begin{equation}
A^0 (x) = \int_{-\infty}^{\infty} d\tau \ a^0 (x,\tau) = \ 
-\frac{e}{4\pi R} \ \frac12 \int_{-\infty}^{\infty} d\tau \ 
\Bigl( \varphi(t - \tau - R/c) + \varphi(t - \tau + R/c) \Bigr) =
-\frac{e}{4\pi R}
\label{eqn:74}
\end{equation}
as required.

A convenient choice of distribution function is
\begin{equation}
\varphi (\alpha) = \frac{1}{2\lambda} e^{-|\alpha|/\lambda} \ ,
\label{eqn:75}
\end{equation}
in which $\lambda$ represents the width of the distribution.  
For this distribution function, the induced potential is given by
\begin{equation}
a^0_\varphi (x,\tau) =  -\frac{e}{4\pi R} \ 
\frac{1}{2\lambda}  \ \frac12
 \ \left[ e^{-\bigl|t - \tau - R/c\bigr|/\lambda} +
 e^{-\bigl|t - \tau + R/c\bigr|/\lambda} \right] \ .
\label{eqn:76}
\end{equation}
We re-write the equations of motion (\ref{eqn:32}) and
(\ref{eqn:33}) as
\begin{eqnarray}
M \ \ddot x^0 &=& -\frac12 \ \frac{e}{c} \ 
\Bigl[ \bigl(\partial_k \tilde{a}^0_\varphi\bigr) \dot x^k -
\frac1c \ (\partial_t - \partial_\tau) \ \tilde{a}^0_\varphi \Bigr]
\label{eqn:32tag}
\\
M \ \ddot x^k &=&  - e \ 
\bigl(\partial_k \tilde{a}^0_\varphi\bigr)\frac{\dot t +1}{2} \ ,
\label{eqn:33tag}
\end{eqnarray}
where the field
\begin{equation}
\tilde{a}^0_\varphi = 2 \lambda  \ a^0_\varphi
= -\frac{e}{4\pi R} \ \frac12
 \ \left[ e^{-\bigl|t - \tau - R/c\bigr|/\lambda} +
 e^{-\bigl|t - \tau + R/c\bigr|/\lambda} \right]
\label{eqn:a-twiddle}
\end{equation}
is defined to include the factor $2 \lambda$ and so has the units of
$A^0$.  In the low energy limit ($v/c \sim 0$), with $t \sim \tau$
(and $\dot t \sim 1$),
we obtain the standard equations of
motion for a particle in a classical Yukawa potential,
\begin{equation}
M \ \ddot x^0 = 0 \qquad
M \ \ddot{\vec{x}} =  e \ (- \nabla \tilde{a}^0_\varphi) \ .
\label{eqn:motion}
\end{equation}
where
\begin{equation}
\tilde{a}^0_\varphi (x,\tau) = -\frac{e}{4\pi R} \ e^{-R/\lambda c}
 \ .
\label{eqn:77}
\end{equation}
Thus, the parameter $\lambda$ may be estimated by the
experimental precision of low energy Coulomb scattering.  Clearly
as $\lambda$ becomes very large, corresponding to a wide
distribution of events, the potential
$\tilde{a}^0_\varphi$ approaches the standard Coulomb potential.

The derivatives of the potential may be found from
(\ref{eqn:a-twiddle})
\begin{eqnarray}
\partial_k \tilde{a}^0_\varphi (x,\tau) &=& \partial_k \left\{
-\frac{e}{4\pi R} \ 
\frac12
 \ \left[ e^{-\bigl|t - \tau - R/c\bigr|/\lambda} +
 e^{-\bigl|t - \tau + R/c\bigr|/\lambda} \right] \right\}
\nonumber \\
&=& \frac{e}{4\pi R^2} \ \hat{x}^k \ \frac12 \ 
\left[ e^{-\bigl|t - \tau - R/c\bigr|/\lambda} +
 e^{-\bigl|t - \tau + R/c\bigr|/\lambda} \right]
\nonumber \\ && \mbox{\qquad} 
 -\frac{e}{4\pi R} \ \frac12 \ 
\partial_k \left[ e^{-\bigl|t - \tau - R/c\bigr|/\lambda} +
 e^{-\bigl|t - \tau + R/c\bigr|/\lambda} \right] \ .
\label{eqn:78}
\end{eqnarray}
Using
\begin{eqnarray}
\partial_\alpha e^{-|\xi(x,\tau)|/\lambda} &=& \frac{d}{d\xi} e^{-|\xi(x,\tau)|/\lambda}
\ \partial_\alpha \xi(x,\tau)
\nonumber \\
&=& \frac{d}{d\xi} \left[ \theta(\xi) \ e^{-\xi(x,\tau)/\lambda}
+ \theta(-\xi) \ e^{+\xi(x,\tau)/\lambda} \right] \ \partial_\alpha \xi(x,\tau)
\nonumber \\
&=& \Biggl\{ \delta(\xi) \ e^{-\xi(x,\tau)/\lambda} - \delta(-\xi) \ 
e^{+\xi(x,\tau)/\lambda}
\nonumber \\ && \mbox{\qquad}
+ \theta(\xi) \ e^{-\xi(x,\tau)/\lambda}
\left(-\frac{1}{\lambda} \right)
+ \theta(-\xi) \ e^{+\xi(x,\tau)/\lambda}
\left(\frac{1}{\lambda} \right) \Biggr\}  \ \partial_\alpha \xi(x,\tau)
\nonumber \\
&=& - \frac{1}{\lambda} \ \epsilon\Bigl(\xi(x,\tau)\Bigl)
e^{-|\xi(x,\tau)|/\lambda} \ \partial_\alpha \xi(x,\tau) \ ,
\label{eqn:79}
\end{eqnarray}
we find
\begin{eqnarray}
\partial_k \tilde{a}^0_\varphi (x,\tau) &=&
  \frac{e}{4\pi R^2} \
\hat{x}^k \ \frac12
 \ \left[ e^{-\bigl|t - \tau - R/c\bigr|/\lambda} +
e^{-\bigl|t - \tau + R/c\bigr|/\lambda} \right]
\nonumber \\ && \mbox{\hspace{-1.2 in}}+
 \frac{e}{4\pi R} \ \hat{x}^k \ \frac12
\Biggl[-\frac{1}{\lambda c}  \ 
\epsilon(t - \tau - R/c) \ 
e^{-\bigl|t - \tau - R/c\bigr|/ \lambda}
+ \frac{1}{\lambda c}  \ 
\epsilon(t - \tau + R/c) \ 
e^{-\bigl|t - \tau + R/c\bigr|/\lambda}
 \Biggr]
\label{eqn:80}
\end{eqnarray}
for the space-derivatives, and
\begin{equation}
\partial_t \tilde{a}^0_\varphi (x,\tau) = -
\frac{1}{\lambda} \ \frac12 \ \frac{e}{4\pi R}
\left[ e^{-\bigl|t - \tau - R/c\bigr|/ \lambda}
\epsilon(t - \tau - R/c) + 
e^{-\bigl|t - \tau + R/c\bigr|/\lambda} \epsilon(t - \tau + R/c) \right]
\label{eqn:81}
\end{equation}
and
\begin{equation}
\partial_\tau a^0_\varphi (x,\tau) = +
\frac{1}{\lambda} \ \frac12 \ \frac{e}{4\pi R}
\epsilon(t - \tau - R/c) \ 
\left[ e^{-\bigl|t - \tau - R/c\bigr|/ \lambda}
+ 
\epsilon(t - \tau + R/c) \ 
e^{-\bigl|t - \tau + R/c\bigr|/\lambda}
\right]
\label{eqn:82}
\end{equation}
for the time-derivatives.
Using equations (\ref{eqn:80}), (\ref{eqn:81}), and (\ref{eqn:82}) in
the equations of motion (\ref{eqn:32tag}) and (\ref{eqn:33tag}), we
arrive at the well-behaved system of equations
\begin{eqnarray}
M \ddot x^k &=& - \frac{e^2}{4\pi R^2} \ \hat{x}^k \ 
\frac{\dot t + 1}{2} \ \times
\nonumber \\&& \mbox{\hspace{-.5 in}}
\frac12\Biggl\{
\Bigl[ 1+\frac{R}{\lambda c} \epsilon(t-\tau  -R/c) \Bigr]
 e^{-\bigl|t - \tau - R/c\bigr|/\lambda} +
\Bigl[ 1-\frac{R}{\lambda c} \epsilon(t-\tau  +R/c) \Bigr]
 e^{-\bigl|t - \tau + R/c\bigr|/\lambda} 
\Biggr\}
\label{eqn:83}
\end{eqnarray}
and
\begin{eqnarray}
M \ddot t &=& -\frac{1}{2c^2} \ \frac{e^2}{4\pi R^2}\frac12
\Biggl\{ \Bigl[
\hat{x}^k \dot x_k (1 + \frac{R}{\lambda c}
\epsilon(t-\tau  -R/c) ) -
\frac{2R}{\lambda} \epsilon(t-\tau  -R/c) \Bigr]
e^{-\bigl|t - \tau - R/c\bigr|/\lambda}
\nonumber \\&&\mbox{\qquad}
+ \Bigl[
\hat{x}^k \dot x_k (1 - \frac{R}{\lambda c}
\epsilon(t-\tau  +R/c) ) -
\frac{2R}{\lambda} \epsilon(t-\tau  +R/c) \Bigr]
e^{-\bigl|t - \tau + R/c\bigr|/\lambda} \Biggr\}
\label{eqn:84}
\end{eqnarray}
for the motion of the test event.
In the low energy limit, the equations of motion simplify to
\begin{eqnarray}
M \ddot x^k &=& - \frac{e^2}{4\pi R^2} \hat{x}^k
\frac12 \ 
\Biggl\{
\Bigl[ 1+\frac{R}{\lambda c} \epsilon(-R/c) \Bigr]
 e^{-R/\lambda c} +
\Bigl[ 1-\frac{R}{\lambda c} \epsilon(+R/c) \Bigr]
 e^{-R/\lambda c} 
\Biggr\}
\nonumber \\
&=& - \frac{e^2}{4\pi R^2} \hat{x}^k
 \ e^{-R/\lambda c} \ 
 \Bigl[ 1-\frac{R}{\lambda c}\Bigr]
\label{eqn:85}
\end{eqnarray}
and
\begin{eqnarray}
M \ddot t &=& - \frac{e^2}{4\pi R^2} \ \frac{1}{2} \ 
\frac{1}{2c^2} \ 
\Biggl\{ \Bigl[
\hat{x}^k \dot x_k (1 + \frac{R}{\lambda c} \epsilon(-R/c) ) -
\frac{2R}{\lambda} \epsilon(-R/c) \Bigr] e^{-R/\lambda c}
\nonumber \\&&\mbox{\qquad}
+ \Bigl[
\hat{x}^k \dot x_k (1 - \frac{R}{\lambda} \epsilon(R/c) ) -
\frac{2R}{\lambda} \epsilon(R/c) \Bigr] e^{-R/\lambda c}\Biggr\}
\nonumber \\
&=& - \frac{e^2}{4\pi R^2} \ \frac{1}{2c^2} \ \hat{x}^k
\dot x_k \  \ e^{-R/\lambda c} \ 
 \Bigl[ 1-\frac{R}{\lambda c}\Bigr]
\nonumber \\
&=& o(v/c)  \ .
\label{eqn:86}
\end{eqnarray}
Numerical solutions for the equations of motion (\ref{eqn:83})
and (\ref{eqn:84}) are shown in Figures 1, 2, and 3.  It may be
seen that the trajectories of the test event are indistinguishable
from the Maxwell case, when $\lambda > 10^{-6}$ seconds,
corresponding to a photon mass $m_\gamma \sim 10^{-9}$ eV.
A more conservative estimate may be found by taking
the accepted experimental error in the
photon mass as an actual mass spectrum for the photon.
Then $m_\gamma \simeq 6 \times 10^{-16}$
eV \cite{pdg}, which corresponds to $\lambda \simeq 1$ second.  

\section{Summary and Conclusions}

In their initial formulation of covariant quantum mechanics,
Horwitz and Piron \cite{H-P} constructed an interacting theory of
events, mediated by standard $\tau$-independent Maxwell gauge
fields.  This construction was later deemed incomplete
\cite{saad,group} because of the self-consistency problem: on the
one hand, particle worldlines are traced out by the evolution
of trajectories $x^\mu(\tau)$ under the local influence of
the Maxwell field $A^\mu(x)$; on the other hand, $A^\mu(x)$
depends upon the entire particle worldline --- it is
induced by the Maxwell current $J^\mu(x)$.  For each event,
the motion would depend on the entire worldline of the other, which
induces the field in which it moves.  There can be no
{\it a priori} guarantee that the
resulting trajectories $x^\mu(\tau)$ will produce the fields
$A^\mu(x)$.  The pre-Maxwell theory, developed to insure a
well-posed theory of interactions, contains
fields $a^\alpha (x,\tau)$ and a time scale $\lambda$ required so
that $\lambda a^\mu$ and $A^\mu = \int d\tau \ a^\mu$ will have
the same units.  Despite the many successes of the relativistic
quantum theory, the $\tau$-dependence of the fields and the
presence of $\lambda$ in the equations of motion lead to
difficulties at the classical level.  Introducing uncertainty in
the $\tau$-synchronization of the interaction, by modelling the source
particle current as
a distribution of event currents along the worldline, eliminates these
problems, and permits comparison with the standard Maxwell theory
at low energy.  When the range of the photon mass spectrum is
taken small enough, then the equations of motion for the
off-shell electromagnetic theory coincide with those of the
standard Maxwell theory, within experimental limits.  The model
thus, provides an initial estimate of $\lambda$: if $\lambda$
is larger than about $10^{-6}$, then off-shell phenomena must be
observable in scattering.

A cut-off in
the photon mass spectrum was also found to be necessary for the
renormalization of off-shell quantum electrodynamics \cite{qed}.
The superposition of currents given in (\ref{eqn:70}) leads to
a similar expression for the fields,
\begin{equation}
a^\mu_\varphi (x,\tau) =
\int_{-\infty}^{\infty} d\alpha \ \varphi(\alpha) \ 
a^\mu (x,\tau-\alpha) \ .
\label{eqn:sup-a}
\end{equation}
From the Fourier expansion \cite{qed}
\begin{equation}
a^\mu (x,\tau) = \sum_{s={\rm polarizations}} \int \frac{d^4 k}
{2\kappa} \left[ \varepsilon^\mu_s a(k,s)
e^{i(k\cdot x + \sigma \kappa \tau)/\hbar} +
\varepsilon^{\mu*}_s a^*(k,s)
e^{-i(k\cdot x + \sigma \kappa \tau)/\hbar} \right]
\label{eqn:5.14}
\end{equation}
where the five-dimensional mass shell condition is
$\kappa = \sqrt{k^2}$, we see that under the convolution
(\ref{eqn:sup-a}), the Fourier transform $a(k,s)$ will
acquire a factor
\begin{equation}
a(k,s) \longrightarrow a(k,s) \int d\alpha \ 
e^{i \kappa \alpha/\hbar} \ 
\varphi(\alpha) \ .
\label{eqn:sup-a2}
\end{equation}
Using (\ref{eqn:75}) for $\varphi(\alpha)$, we find this factor to
be
\begin{equation}
\int d\alpha \ e^{i \kappa \alpha/\hbar} \varphi(\alpha) =
\int d\alpha \ e^{i \kappa \alpha/\hbar} \ 
\frac{1}{2\lambda} \ e^{-|\alpha|/\lambda} =
\frac{1}{1+ (\lambda \kappa / \hbar)^2} \ ,
\label{eqn:cutoff}
\end{equation}
which provides an adequate cut-off for the renormalization of
off-shell quantum electrodynamics.

Shnerb and Horwitz \cite{shnerb} have given an interpretation of
$\hbar/\lambda c^2$ as the width of the mass distribution of the
off-shell photons, yielding $\lambda c$ as a coherence length for
the photon-matter interaction.  The model offered above may be
seen as extending this argument to the classical level, as well
as providing a mechanism for the narrow width of the photon mass
spectrum, based on the structure of matter currents.  By using
$\lambda$ to characterize the uncertainty in the
$\tau$-synchronization of the interacting worldlines, we provide
a smooth transition between off-shell formalism and the standard
Maxwell theory.  In Maxwell theory, interactions, characterized by
$J^\mu (x) A_\mu (x)$, take place between entire worldlines, and
the massless photon permits no mass exchange between the
interacting particles.  In the off-shell formalism, interactions
take place between the events which trace out worldlines,
through the exchange of massive $\tau$-dependent photons which may
carry mass
between particles.  However, as $\lambda$ becomes large in this
model, the mass spectrum of the off-shell photon and mass exchange
become more restricted, the $\tau$-dependence of the fields
becomes less significant, and the test event interacts,
effectively, with a larger section of the source worldline.  In
the limit that $\lambda \rightarrow 0$, the model goes over to
the standard Maxwell theory in a proper-time formulation.

For $\lambda$ large but finite, the pre-Maxwell theory remains
well-posed, and the expected off-shell phenomena may be compared
with the experimental limit.  Such a situation suggests that
evidence for the pre-Maxwell theory may be found in small
deviations from standard electrodynamics.

%
%

\end{document}